\documentclass[conference]{IEEEtran}
\usepackage[utf8]{inputenc}
\usepackage[T1]{fontenc}

\usepackage{graphicx}
\usepackage{subfigure}
\usepackage{caption}
\usepackage{float}
\usepackage{balance}
\usepackage{adjustbox}
\usepackage{multirow}
\usepackage{listings}
\usepackage[newfloat]{minted}
\usepackage{paralist}
\usepackage[inline, shortlabels]{enumitem}
\usepackage{algorithm} 

\usepackage{xurl}
\usepackage[hidelinks]{hyperref}
\hypersetup{colorlinks=true,linkcolor=blue,citecolor=blue,urlcolor=blue}
\usepackage{nth}
\usepackage{soul}
\usepackage[normalem]{ulem}
\usepackage[inline]{enumitem}

\usepackage{amsmath}
\usepackage{array,tabularx,booktabs,makecell,colortbl, multirow}

\usepackage{pifont}
\usepackage{fontawesome5}
\newcommand{\cmark}{\ding{51}}
\newcommand{\xmark}{\ding{55}}

\usepackage{xspace}
\newcommand{\subhead}[1]{\medskip \noindent{\textbf{#1:}}}
\newcommand{\eg}{\textit{e.g.,}\xspace}
\newcommand{\ie}{\textit{i.e.,}\xspace}
\newcommand{\mitre}{MITRE ATT\&CK\xspace}
\newcommand{\etal}{\textit{et al.}\xspace}
\newcommand{\sol}{\textsc{Ahlert}\xspace}

\usepackage{color, xcolor} 
\usepackage[most]{tcolorbox}

\usepackage{tikz}
\usetikzlibrary{positioning,arrows.meta,fit,backgrounds,calc,shapes.geometric}
\definecolor{beaublue}{rgb}{0, 0, 0}
\definecolor{bluegray}{rgb}{0, 0, 0}
\newcommand{\circled}[1]{\tikz[baseline=(myanchor.base)]      \node[circle,fill=beaublue,draw=bluegray,text=white,inner sep=1.5pt] (myanchor) {\footnotesize #1};}

\newcommand{\circledcolor}[2]{\tikz[baseline=(myanchor.base)] \node[circle,fill=#1,draw=#1,text=white,inner sep=1.5pt] (myanchor) {\footnotesize #2};}
\definecolor{circlegray}{RGB}{160,160,160}
\definecolor{circleblue}{RGB}{0,32,96}
\definecolor{circleorange}{RGB}{204,85,0}
\newcommand{\circledgray}[1]{\circledcolor{circlegray}{#1}}
\newcommand{\circledblue}[1]{\circledcolor{circleblue}{#1}}
\newcommand{\circledorange}[1]{\circledcolor{circleorange}{#1}}

\definecolor{reportblue}{RGB}{0,32,96}
\definecolor{kggreen}{RGB}{73,119,38}
\definecolor{validatedorange}{RGB}{198,89,9}

\definecolor{challengecolor}{rgb}{0.8, 0.2, 0.2}
\newcommand{\challengecircled}[1]{\tikz[baseline=(myanchor.base)] 
\node[circle,fill=challengecolor,draw=challengecolor,text=white,inner sep=1pt] 
(myanchor) {\footnotesize \textsc{#1}};}

\definecolor{innovationcolor}{rgb}{0.6, 0.2, 0.8}

\definecolor{rqtagcolor}{RGB}{70,117,188} 

\newlength{\rqtagwidth}
\newlength{\rqtaghalf}

\newcommand{\akash}{\color{black}}

\definecolor{Baseline1}{RGB}{255,150,100} 
\definecolor{Baseline2}{RGB}{150,200,255} 
\definecolor{Baseline3}{RGB}{80,80,80} 
\definecolor{Baseline4}{RGB}{153,88,42} 
\definecolor{Baseline5}{RGB}{42,157,143} 
\definecolor{Baseline6}{RGB}{167,201,87} 
\definecolor{Baseline7}{RGB}{156,117,193}
\definecolor{Baseline8}{RGB}{233,196,106}
\definecolor{Baseline9}{RGB}{100,181,246}
\definecolor{Baseline10}{RGB}{129,199,132}

\newcommand{\sys}[2]{\textcolor{#1}{\rule{0.8em}{0.8em}}\hspace{0.3em}#2}

\newcommand{\blone}{\sys{Baseline1}{\textsc{\sol-ChatGPT}\xspace}}

\newcommand{\blseven}{\sys{Baseline7}{\textsc{\sol-Qwen}\xspace}}
\newcommand{\bleight}{\sys{Baseline8}{\textsc{\sol-Foundation-Sec}\xspace}}
\newcommand{\blfour}{\sys{Baseline4}{\textsc{ChatGPT-ots}\xspace}}

\newcommand{\blnine}{\sys{Baseline9}{\textsc{Qwen-ots}\xspace}}
\newcommand{\blten}{\sys{Baseline10}{\textsc{Foundation-Sec-ots}\xspace}}

\definecolor{GreenColor}{HTML}{E8F5E9}
\definecolor{GrayColor}{HTML}{F5F5F5}
\definecolor{RedColor}{HTML}{FDECEA}
\definecolor{YellowColor}{HTML}{FFF8E1}
\newcommand{\greencell}[1]{\cellcolor{GreenColor}\textcolor{green!50!black}{#1}}

\newcommand{\redcell}[1]{\cellcolor{RedColor}\textcolor{red!60!black}{#1}}

\begin{document}
\pagestyle{empty}
\thispagestyle{empty}
\bstctlcite{IEEEexample:BSTcontrol} 

\title{Evidence-Grounded Retrieval for Investigation Hunt Lead Generation from CTI Reports}

\author{ 
    \IEEEauthorblockN{
    Akash Prakash\IEEEauthorrefmark{5},
    Boubakr Nour\IEEEauthorrefmark{1},
    Makan Pourzandi\IEEEauthorrefmark{1},
    Chadi Assi\IEEEauthorrefmark{5}, and
    Mourad Debbabi\IEEEauthorrefmark{5}
    }

    \IEEEauthorblockA{\IEEEauthorrefmark{5}Concordia University, Canada \qquad 
    \IEEEauthorrefmark{1}Ericsson Security Research, Canada}
}
\maketitle

\begin{abstract}
    Threat hunting increasingly depends on converting unstructured knowledge (\eg Cyber Threat Intelligence reports) into actionable \emph{hunt leads}: concise, investigable hypotheses grounded in observable artifacts and adversary techniques. Producing such leads manually is a tedious and hard-to-scale task. Existing automated approaches stop at the entity layer, ignore the defender's operational environment, and analyze each report in isolation. 
    To address these gaps, we {\akash introduce} \sol, a system that automatically extracts relevant, environment-aware, and hunt leads from threat reports through
    \begin{enumerate*}[(i)]
        \item a hybrid retriever that combines dense vector search with multi-hop traversal over a knowledge graph seeded with MITRE ATT\&CK;
        \item an ontology-grounding retrieval augmented generation method that constrains each lead to the defender's own assets and controls; and
        \item an LLM-agnostic framework that emits structured, directly actionable leads rather than loose indicators of compromise.
    \end{enumerate*}
    We evaluate \sol on public CTI reports for well-known APT across multiple proprietary and open-weight models. 
    Hybrid evidence retrieval with ontology grounding raises mean F1 by $\approx2\times$ (0.44 to 0.85) over a single-route flat-RAG baseline, and \sol attains the highest effectiveness score ($\approx$86.95\%) compared with off-the-shelf LLM models.

    {\small \normalfont \noindent \faYoutube \sol~Demo: \url{https://youtu.be/zdCquNNV6sA}}
\end{abstract}


\IEEEpeerreviewmaketitle

\section{Introduction}
\label{sec:intro}
The cyber threat landscape is expanding in volume, velocity, and sophistication faster than security operations teams can keep pace. Contemporary adversaries combine zero-day exploits, living-off-the-land techniques, and supply-chain compromise into campaigns that evade signature-based controls, leaving residual risk that requires human-in-the-loop investigation~\cite{crowdstrike2024global}. \emph{Threat hunting} has emerged to address this gap~\cite{nour2023survey}, with analysts iteratively forming hypotheses about adversary behavior and searching enterprise telemetry for corroborating evidence. 
At the core of this process is a knowledge-transformation challenge: raw, unstructured Cyber Threat Intelligence (CTI)~\cite{strom2018mitre}, including vendor reports,
malware analyses, and incident post-mortems, must be distilled into actionable \emph{hunt leads}. 
CrowdStrike~\cite{crowdstrike2023behind} defines a \emph{hunt lead} as a highly specific, low-fidelity indicator, data point, or anomalous behavior that a hunter identifies and investigates; while not malicious on their own, such leads provide context to trace adversary activity before automated systems alert.

A hunt lead is a combination of: a natural-language hypothesis tied to the supporting evidence and adversary technique it rests on, a triage assessment of its \textsc{severity}, \textsc{priority}, and likely \textsc{impact}, and the concrete \textsc{metrics} and \textsc{artifacts}, such as hosts, processes, and indicators, that the hunter can immediately pivot on.

\subhead{Motivation}
While CTI is widely accessible, constructing actionable hunt leads remains a major operational bottleneck~\cite{maxam2024interview}. A single report can span dozens of pages mixing indicators, \mitre tactics, techniques, and procedures (TTPs), malware details, and narrative; hunters must manually extract observations and map them to the assets and controls of their environment, a process that is slow, error-prone, and hard to scale.
Prior work~\cite{lekssays2025azerg, kim2026anchor, cheng2025ctinexus} improves entity extraction and technique mapping but typically \emph{stops at the entity layer}~\cite{buchel2025sok}, before generating the behavioral hypotheses that actually produce a hunt lead. This is a critical gap, since TTPs are more stable and valuable for defense than atomic IoCs, which adversaries easily change.

\begin{figure*}[t]
    \centering
    \includegraphics[width=\linewidth]{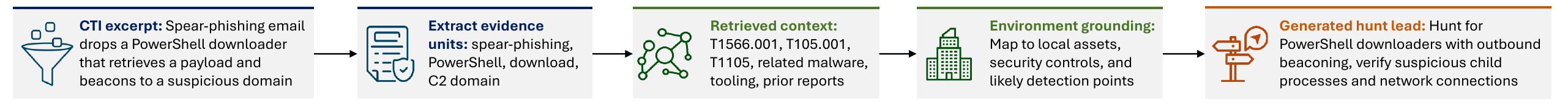}
    \caption{High-level overview of hunt lead extraction.}
    \label{fig:overview-simple}
\end{figure*}

\subhead{Challenges}
Automating hunt lead extraction is hard because a lead is useful only if it is at once faithful to the source intelligence, aware of the defender's environment, and informed by prior knowledge. Three challenges follow:

\noindent \challengecircled{1} \textit{Faithfulness to evidence:} an extracted lead must rest on indicators and techniques that genuinely appear in the source intelligence; fabricated hashes, spurious technique identifiers, or unsupported behavioral claims misdirect a hunt and erode {\akash analyst's} trust;
\noindent \challengecircled{2} \textit{Environment blindness:} a technically valid lead is operationally ineffective if it refers to sensors, controls, or data sources that are absent from the defender's environment; extraction must therefore be grounded in what the defender can actually observe;
\noindent \challengecircled{3} \textit{Absence of relational memory:} each report is typically analyzed in isolation, ignoring the broader graph of actor $\rightarrow$ tool $\rightarrow$ technique $\rightarrow$ asset relationships that experienced hunters use to connect new observations to months or years of prior intelligence.

\subhead{Limitations of existing solutions}
Existing approaches to hunt lead extraction fall short on three fronts:
\circled{1} Most solutions~\cite{lekssays2025azerg, kim2026anchor, cheng2025ctinexus} \emph{stop at the entity layer}: they extract indicators, map techniques, and tag actors, but leave the synthesis of an investigable hypothesis, the lead itself, to the human analyst;
\circled{2} The evidence for a single lead is rarely co-located; it must be assembled across a malware analysis, a technique description, and an asset inventory connected only through a shared actor, tool, or technique, a relational structure that flat keyword or similarity matching cannot stitch together; and
\circled{3} Extracted leads are seldom validated against the defender's own environment, so they may reference sensors, controls, or data sources that the SOC does not actually operate, yielding leads that read well but cannot be hunted.

A practical solution must therefore \emph{unify} flat retrieval, graph traversal, and controlled generation, and be \emph{environment-aware}, filtering leads against a formal description of the defender's infrastructure so only operationally actionable hypotheses reach the analyst. 
As shown in Table~\ref{tab:sota_comparison}, existing solutions such as \textsc{Azerg}~\cite{lekssays2025azerg}, \textsc{Ctinexus}~\cite{cheng2025ctinexus}, and \textsc{TechniqueRAG}~\cite{lekssays2025techniquerag} rarely incorporate this constraint, assuming an abstract defender with a complete sensor fleet, whereas real SOCs cover only a bounded subset of assets, controls, and data sources.

\begin{table}[!t]
    \setlength{\tabcolsep}{2pt}
    \centering
    \caption{Capability comparison between \sol and representative state-of-the-art extraction solutions.
    }
    \label{tab:sota_comparison}
    \renewcommand{\arraystretch}{1.15}
    \resizebox{\linewidth}{!}{%
    \begin{tabular}{rccccc}
        \hline
        \textbf{~}
        & \textbf{Graph}
        & \textbf{KB}
        & \textbf{Environment}
        & \textbf{hunt lead}
        & \textbf{Multi-LLM}\\
        & \textbf{Retrieval}
        & \textbf{Pre-built}
        & \textbf{Grounding}
        & \textbf{Output}
        & \textbf{Evaluation}\\
        \hline
        \textsc{Azerg}~\cite{lekssays2025azerg}
            & \redcell{\xmark} & \redcell{\xmark}
            & \redcell{\xmark} & \redcell{\xmark}
            & \redcell{\xmark} \\
        \textsc{Anchor}~\cite{kim2026anchor}
            & \redcell{\xmark} & \redcell{\xmark}
            & \redcell{\xmark} & \redcell{\xmark}
            & \redcell{\xmark} \\
        \textsc{Ctinexus}~\cite{cheng2025ctinexus}
            & \redcell{\xmark} & \redcell{\xmark}
            & \redcell{\xmark} & \redcell{\xmark}
            & \redcell{\xmark} \\
        \textsc{TechniqueRAG}~\cite{lekssays2025techniquerag}
            & \redcell{\xmark} & \redcell{\xmark}
            & \redcell{\xmark} & \redcell{\xmark}
            & \greencell{\cmark} \\
        \textsc{AgCyRAG}~\cite{kurniawan2025agcyrag}
            & \greencell{\cmark} & \redcell{\xmark}
            & \redcell{\xmark} & \redcell{\xmark}
            & \redcell{\xmark} \\
        \textsc{Beyond RAG}~\cite{hamzic2025beyondrag}
            & \greencell{\cmark} & \greencell{\cmark}
            & \redcell{\xmark} & \redcell{\xmark}
            & \greencell{\cmark} \\
        \sol
            & \greencell{\cmark} & \greencell{\cmark}
            & \greencell{\cmark} & \greencell{\cmark}
            & \greencell{\cmark} \\
        \hline
    \end{tabular}}
\end{table}

\begin{figure*}[t]
    \centering
    \includegraphics[width=.9\linewidth]{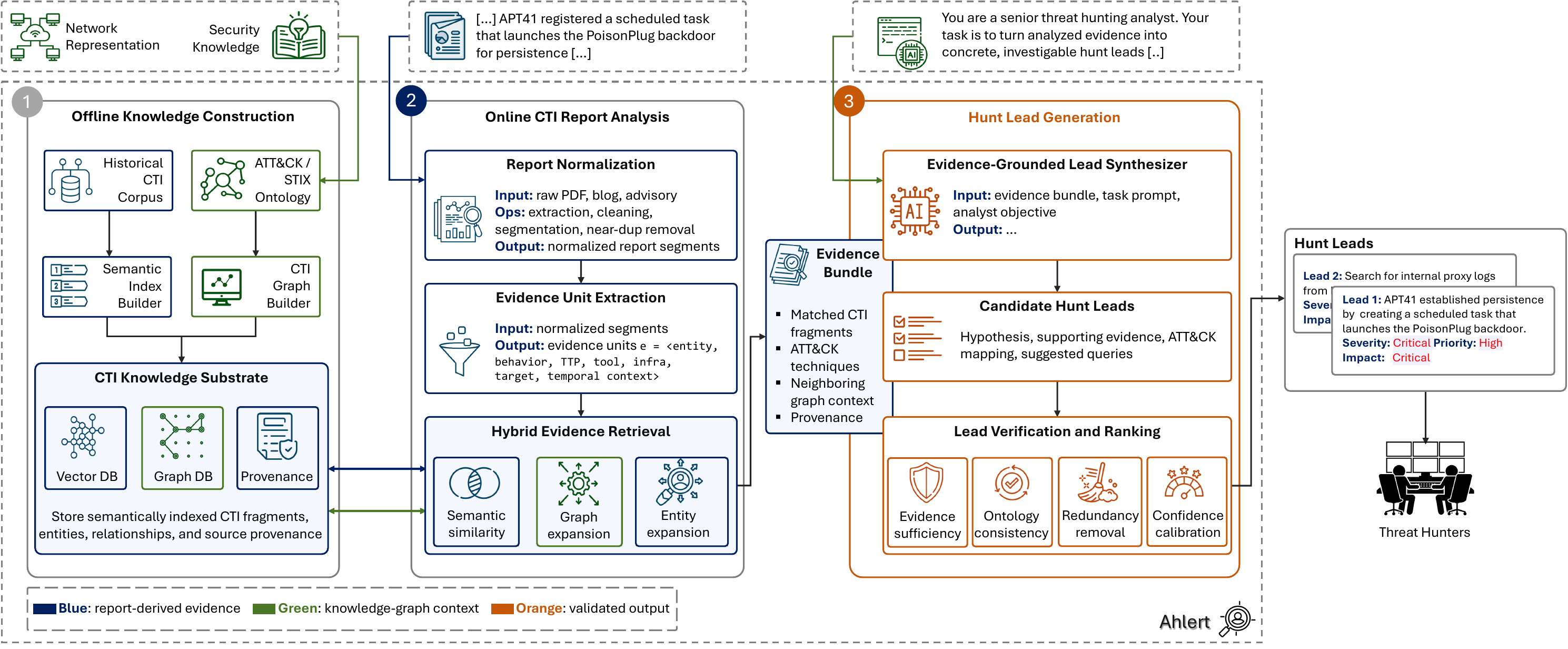}
    \caption{End-to-end architecture of \sol. {\footnotesize Color coding distinguishes \textcolor{reportblue}{report-derived evidence}, \textcolor{kggreen}{knowledge-base context}, and \textcolor{validatedorange}{validated output}.}}
    \label{fig:arch}
\end{figure*}

\subhead{Contributions}
To address these challenges, we design \sol, an automated solution that turns a threat report into actionable hunt leads by extracting evidence, enriching it through hybrid evidence retrieval,
grounding it to the defender's environment, and synthesizing prioritized hypotheses (Fig.~\ref{fig:overview-simple}). The main contributions are:

\begin{itemize}[leftmargin=*, noitemsep, topsep=2pt]
    \item {\akash We introduce} \sol, a hybrid retrieval-and-generation system for automated hunt lead extraction from unstructured threat reports that combines dense vector retrieval with cyber knowledge graph reasoning to enrich report-derived evidence through explicit relationships among techniques, tools, actors, campaigns, assets, and indicators.  

    \item We introduce an ontology-aware and post-filtering mechanism that constrains LLM outputs to the defender's asset and control universe, thereby reducing environment-irrelevant leads.

    \item {\akash We present} an evidence-grounded generation stage supported by LLM that synthesizes candidate hunt leads from the retrieved evidence and then verifies and ranks them, so that \sol emits fully phrased, prioritized leads rather than loose indicators of compromise.

    \item We evaluate \sol across multiple contemporary LLMs, including proprietary and open-weight models, on public CTI reports and knowledge base.
    Obtained results show that \sol consistently improves hunt lead quality,
    raising hunt lead F1 score by $\approx2\times$ (0.44 to 0.85) over a single-route flat-RAG baseline and
    attaining a  highest effectiveness score (avg.~86.95\%).

\end{itemize}

\section{\sol: \underline{A}utomated \underline{H}unt \underline{L}ead \underline{E}xtraction using \underline{R}etrieval-augmented \underline{T}hreat intelligence}
\label{sec:sol}

\subsection{System Overview}
\label{subsec:overview}

\subhead{Overview}
\sol is a hybrid-retrieval, ontology-aware system that transforms an unstructured threat report into a ranked list of environment-consistent hunt leads. Unlike single-route RAG~\cite{edge2024graphrag} that conditions the generator on one modality alone, \sol fuses three complementary routes, semantic similarity, graph expansion, and entity expansion, into a unified evidence bundle grounded in a defender environment ontology. 
Fig.~\ref{fig:arch} illustrates the high-level overview of \sol and its working principle.
\sol is organized into three phases that move a report from raw text to validated hunt leads.

The first phase is \textit{Offline Knowledge Construction} (Step~\circledgray{1}), which is prepared in advance.
A historical CTI corpus and a curated cyber knowledge base are ingested into a persistent CTI knowledge substrate that exposes both a semantic index and a relationship graph, together with the provenance of every stored item. 
The second phase, \textit{Online CTI Report Analysis} (Step~\circledblue{2}), runs when an analyst submits a report. It normalizes the raw input into clean segments, extracts structured evidence units capturing the entities, behaviors, techniques, tools, infrastructure, targets, and temporal context of each passage, and enriches them against the substrate through semantic similarity, graph expansion, and entity expansion into a bundle of matched fragments, associated techniques, neighboring graph context, and provenance. 
The third phase, \textit{Hunt Lead Generation} (Step~\circledorange{3}), is driven by an analyst task prompt. An evidence-grounded synthesizer drafts candidate leads, which are then verified and ranked for evidence sufficiency, ontology consistency, redundancy, and confidence before the final ranked leads are delivered to the threat hunter. 
\sol emits structured, evidence-backed leads, each carrying its ATT\&CK mapping and source provenance, that a threat hunter can execute directly against enterprise telemetry.

\subhead{Novelty}
The design of \sol introduces three novelties that distinguish it from existing CTI solutions:
\begin{enumerate*}
    \item[] \challengecircled{1} \textit{Hybrid evidence retrieval:} 
    the retrieval stage fuses three complementary routes, semantic similarity, graph expansion, and entity expansion, so report-derived evidence reaches the generator through topical, structural, and relational paths rather than text matching alone;

    \item[] \challengecircled{2} \textit{Ontology-aware grounding:} 
    when the defender's environment is provided, \sol constrains leads to reference it and prunes any that do not, keeping emitted leads operationally actionable rather than generic; and

    \item[] \challengecircled{3} \textit{Model-agnostic generation:} 
    the generation stage is decoupled from retrieval behind a uniform interface, so the same framework runs unchanged across proprietary and open-weight models, enabling a fair comparison across model families.

\end{enumerate*}

\subsection{Offline Knowledge Construction}
\label{subsec:offline}
Ahead of any analysis, \sol assembles a persistent \emph{CTI knowledge substrate} as its long-term knowledge. A historical CTI corpus and a curated cyber knowledge base (KB) are ingested once and organized along two complementary views: a \emph{semantic index}~\cite{wang2020minilm, johnson2021faiss} capturing the meaning of CTI fragments for similarity-based recall, and a \emph{relationship graph}~\cite{strom2018mitre, mitre-tie} recording entities and the actor, tool, technique, and asset links among them. {\akash The substrate stores semantically indexed CTI fragments, their entities and relationships, and the source provenance of each item, so that evidence surfaced later can be traced back to where it came from.}

\subhead{CTI Graph Builder}
\sol's CTI graph is the structural backbone of the retrieval phase. It records cybersecurity entities, such as threat groups, techniques, software, and the reports that describe them, together with the relationships among them, drawn from community-curated sources such as MITRE ATT\&CK and a corpus of historical incident reports. 
The ATT\&CK/TIE backbone is loaded offline, while each newly uploaded CTI report is incrementally merged into the knowledge graph during ingestion, so that retrieval can connect new observations to prior intelligence and reach an actor's known techniques in only a couple of hops.

\subhead{Semantic Index Construction}
Alongside the graph, the same corpus is embedded into a dense vector index so that passages can later be recalled by meaning rather than by exact keywords. 
Each document is segmented into overlapping passages of 600 tokens with an 80-token overlap, preserving cross-sentence context while keeping each unit small enough to embed precisely. Every passage is encoded with the \texttt{all-MiniLM-L6-v2} sentence transformer~\cite{wang2020minilm} into a $384$-dimensional vector, $L_2$-normalized so that inner-product search equals cosine similarity. The collection is indexed with a FAISS HNSW (hierarchical navigable small-world) graph index~\cite{johnson2021faiss, malkov2020hnsw}, giving logarithmic-time approximate nearest-neighbor search at high recall. Each vector is stored with lightweight metadata (source document, passage offset, and provenance), so a retrieved passage can be traced to its origin and fused with graph-derived evidence at query time.

\subsection{Online CTI Report Analysis}
\label{subsec:online}
For a CTI report, \sol processes it online in three steps:
\begin{enumerate*}[(i)]
    \item report normalization takes the raw/unstructured input and applies extraction, cleaning, segmentation, and near-duplicate removal to yield clean, normalized report segments;
    \item evidence unit extraction then converts those segments into structured evidence units, each capturing the salient facets of a passage, namely the entity, behavior, technique, tool, infrastructure, target, and temporal context it describes; and 
    \item hybrid evidence retrieval enriches these units against the offline substrate through three parallel routes, semantic similarity, graph expansion, and entity expansion, so that report-derived evidence is augmented with related knowledge-base context. 
\end{enumerate*}
The result is a consolidated \emph{evidence bundle} comprising the matched CTI fragments, the associated techniques, the neighboring graph context, and the provenance of each element.

\subhead{Hybrid Evidence Retrieval}
Each evidence unit is enriched against the offline knowledge substrate through three complementary retrieval routes.
\begin{enumerate*}[(i)]
    \item \emph{Semantic similarity} recalls report passages and prior CTI that are close in meaning to the unit,
    \item \emph{graph expansion} follows relationships in the cyber-knowledge graph to bring in associated techniques, tooling, and threat groups; and
    \item \emph{entity expansion} pulls in connected entities such as actors, assets, and indicators. 

\end{enumerate*}

\begin{figure}[!t]
    \centering
    \includegraphics[width=\linewidth]{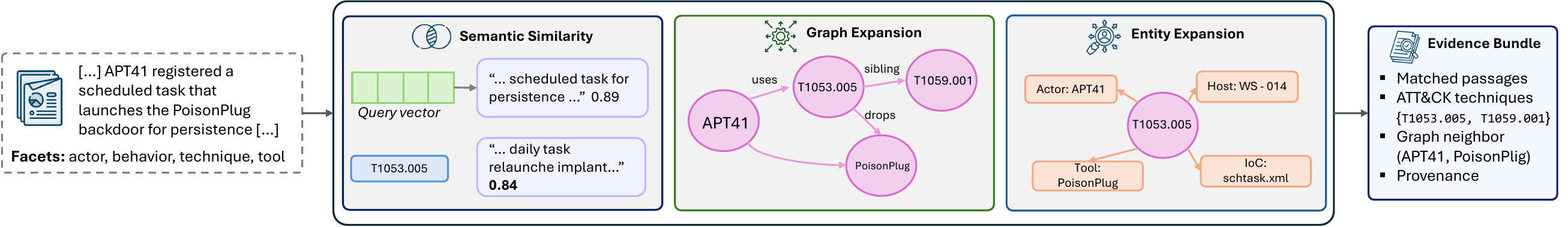}
    \caption{Illustrative example of hybrid evidence retrieval.}
    \label{fig:retrieval-example}
    \vspace{-0.3cm}
\end{figure}

Fig.~\ref{fig:retrieval-example} walks a single evidence unit through the three routes with a concrete example. Starting from unstructured input, \emph{semantic similarity} embeds the unit and runs approximate-nearest-neighbor search over the semantic index, recalling the closest prior passages (\eg at cosine similarity $0.89$ and $0.84$) and mapping them onto technique \texttt{T1053.005}; \emph{graph expansion} traverses the relationship graph from that technique to the actor \textsc{APT41} and to sibling techniques and tooling it is linked to (\eg \texttt{T1059.001}); and \emph{entity expansion} pulls in the concrete entities connected to those nodes, such as the implant \textsc{PoisonPlug}, the affected host, and the scheduled-task indicator. Each bundle is deliberately compact and quality-controlled: the top $k{=}10$ passages survive a $0.15$ minimum-relevance cutoff and optional cross-encoder re-ranking (\texttt{ms-marco-MiniLM-L-6-v2}); graph expansion is capped at $20$ technique and $10$ group candidates linked to the report's focus actor, and every retained element carries source provenance.

\subsection{Hunt Lead Generation}
The final phase of \sol turns the evidence bundle assembled by retrieval into a ranked list of \textit{hunt leads}: concise, action-oriented hypotheses that a threat hunter can execute directly against enterprise telemetry.
A hunt lead is not a restatement of the report but an \textit{investigable hypothesis} about what to look for and where, grounded in the defender's own environment whenever a description of it is available. 
As shown in Fig.~\ref{fig:arch}, generation proceeds in two steps guided by the analyst's task prompt and objective: an evidence-grounded synthesizer that first drafts candidate hunt leads from the evidence bundle, and these candidates are then verified and ranked so that only validated leads reach the threat hunter.

\subhead{Evidence-Grounded Lead Synthesis}
The synthesizer consumes the evidence bundle produced by retrieval, comprising the matched report fragments, their associated ATT\&CK techniques, the neighboring graph context, and the provenance of each item, together with the analyst's task prompt and objective (see Appendix~\ref{app:prompt}).
From this grounded context, \sol drafts a set of candidate hunt leads, each pairing a behavioral hypothesis with the supporting evidence it rests on, the relevant technique mapping, and concrete queries the hunter can run. Because every candidate is tied back to retrieved evidence, the synthesizer is steered towards investigable hypotheses rather than a paraphrase of the report.

\subhead{Lead Verification and Ranking}
The candidate leads pass through a verification and ranking step before they reach the threat hunter. Each lead is checked for:
\begin{enumerate*}[(i)]
    \item evidence sufficiency: so that unsupported hypotheses are discarded a lead is evidence sufficient if at least one retrieved evidence unit semantically supports it above a confidence threshold; 
    \item consistency with the defender's environment: so that operationally infeasible leads are pruned, retaining the unfiltered set if this would discard every candidate; 
    \item redundancy: so that near-duplicate leads are merged; and
    \item calibrated confidence score and ranked to ensure that what reaches the threat hunter is both grounded in the report and relevant to the defender's estate. 
\end{enumerate*}

\begin{figure}[!t]
    \centering
    \includegraphics[width=.8\columnwidth]{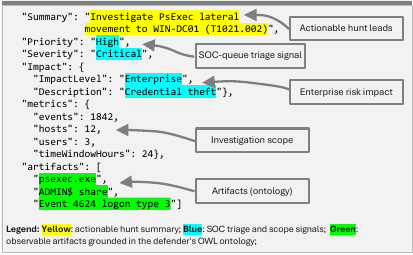}
    \caption{Excerpt of generated hunt lead by \sol.}
    \label{fig:Metrices}
    \vspace{-0.3cm}
\end{figure}

Ranking is performed by the same generator in a single verification pass over the candidate set, guided by the task prompt (Appendix~\ref{app:prompt}): each surviving lead is assigned a calibrated confidence score that reflects the strength and amount of supporting evidence, its consistency with the defender's environment, and the assessed severity of the implied activity. Leads are then ordered by this score, with severity breaking ties, so the highest-confidence, most operationally relevant hypotheses surface first.
Each validated lead is emitted in the fixed schema of Fig.~\ref{fig:Metrices}.

\section{Experimental Setup}
We built a PoC of \sol in Python 3.12\footnote{\faYoutube~\sol's Demo: \url{https://youtu.be/zdCquNNV6sA}}. 
The knowledge base is maintained in Neo4j with a FAISS-HNSW index, and semantic matching uses \texttt{all-MiniLM-L6-v2}~\cite{wang2020minilm} embeddings. Decoding uses temperature $0$ for GPT-4.1-mini and nucleus sampling for the open-weight models (Qwen-2.5-7B, $T{=}0.7$; Foundation-Sec-8B, $T{=}0.3$; both $\mathrm{top}\text{-}p{=}0.9$).
The experiments were performed on a VM running Ubuntu 20.04 LTS (Linux kernel 5.4), provisioned with 30 vCPUs on an AMD EPYC 7702 processor, 211 GB of RAM, and a single NVIDIA A100, SXM4 GPU with 80 GB of memory.

\subhead{Knowledge Base}
The Knowledge Base merges two datasets that are
highly complementary:
(i) \textsc{Enterprise ATT\&CK}~\cite{strom2018mitre}, which provides the canonical taxonomy of adversarial tradecraft: attack patterns (techniques and sub-techniques), intrusion sets, malware, tools, and the \textit{uses} relationships among them; and
(ii) MITRE \textsc{Technique Inference Engine (Tie)}~\cite{mitre-tie}, which contains several thousand historical incident reports annotated with the techniques, threat actor groups, software, and campaigns that each report describes.
ATT\&CK supplies the \textit{what} of attacker behavior while \textsc{Tie} supplies the \textit{who} and the \textit{where-we-have-seen-this}.

\subhead{Baselines}
{\akash As no directly comparable prior work exists, we evaluate \sol against six configurations spanning three \sol-augmented generators and their corresponding off-the-shelf (\textsc{Ots}) counterparts:}
\begin{enumerate*}[(1)]
    \item \blone:   \sol-augmented ChatGPT\footnote{GPT-4.1-mini: \url{https://platform.openai.com/docs/models/gpt-4.1-mini}},
    \item \blseven: \sol-augmented Qwen\footnote{Qwen-2.5-7B-Instruct: \url{https://huggingface.co/Qwen/Qwen2.5-7B-Instruct}},
    \item \bleight: \sol-augmented Cisco Foundation-Sec\footnote{Foundation-Sec-8B: \url{https://huggingface.co/fdtn-ai/Foundation-Sec-8B}},
    \item \blfour,
    \item \blnine, and
    \item \blten.
\end{enumerate*}

\subhead{Threat Reports}
We used the APTNotes repository\footnote{APTNotes repository: \url{https://github.com/aptnotes/data/}}, a collection of public threat intelligence reports from vendors including Mandiant, Trend Micro, and CrowdStrike, and selected four reports on APT41.

\subhead{Ground truth}
A domain expert curated a ground truth set of investigable hunt leads spanning the intrusion's initial-access, execution, persistence, and exfiltration phases for the associated threat.
Each ground truth lead is a single imperative sentence following the same schema as \sol (Fig.~\ref{fig:Metrices}).

\subhead{Ontologies}
The ontologies used in this work were developed in Web Ontology Language (OWL) and constructed using Protégé\footnote{Protégé 5.6: \url{https://protege.stanford.edu/}}.
The cybersecurity ontology~\cite{syed2016uco} contains 12,838 triples and models over 4,280 unique entities, including 103 APT groups, 296 malware families, and 2,896 CVE vulnerabilities, interconnected through 115 relationship types.

{\akash The system ontology models 79 node instances across six component types}: 22 virtual machines, 14 servers, 13 firewalls, 10 applications, 9 services, and 11 workstations.

\section{Evaluation}
\label{sec:eval}

\smallskip
\noindent \textbf{RQ1 - \textit{Lead Relevance}: How relevant are the generated hunt leads to the given threat report and the ontology?}
\noindent To address \textsc{RQ1}, we measured the lead relevance by computing the semantic similarity between each generated hunt lead and a curated set of ground truth leads.
We encoded the \texttt{summary} field of every lead using \textit{all-MiniLM-L6-v2}~\cite{wang2020minilm}.
We consider a true positive lead each generated lead whose maximum semantic similarity to any ground truth lead meets or exceeds the threshold $\tau_{\mathrm{gt}}$. Matching uses independent maximum-similarity checks for generated and ground truth leads rather than a one-to-one assignment.
We set \mbox{$\tau_{\mathrm{gt}}=0.82$} to balance strictness
with tolerance (see RQ4).

\begin{figure}[!t]
    \centering
    \subfigure[Precision.]{\includegraphics[width=0.32\linewidth]{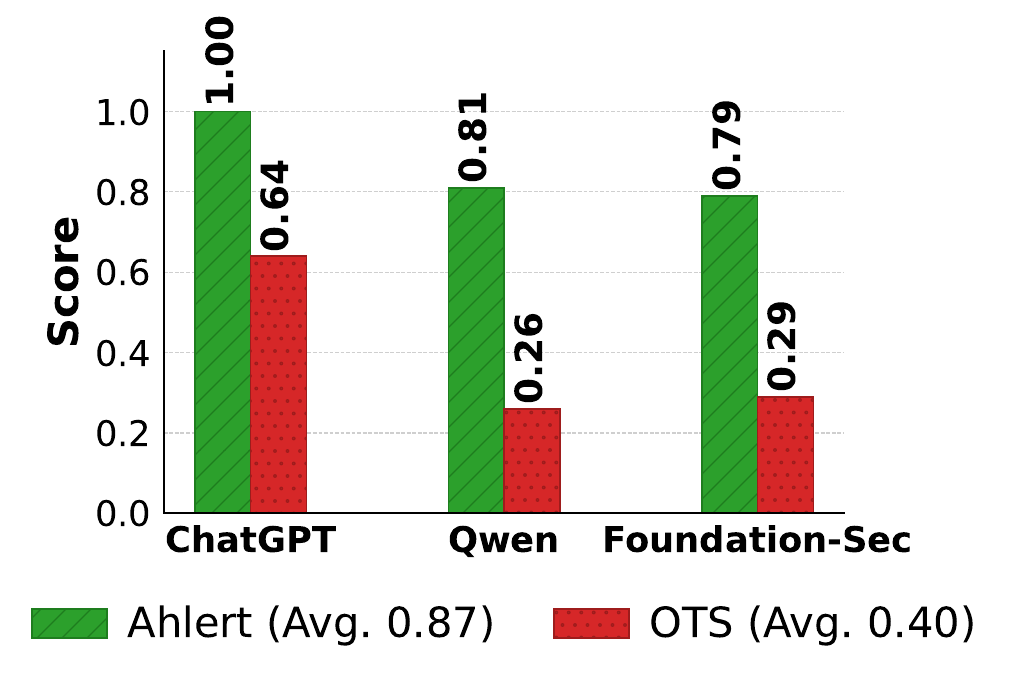}}
    \subfigure[Recall.]{\includegraphics[width=0.32\linewidth]{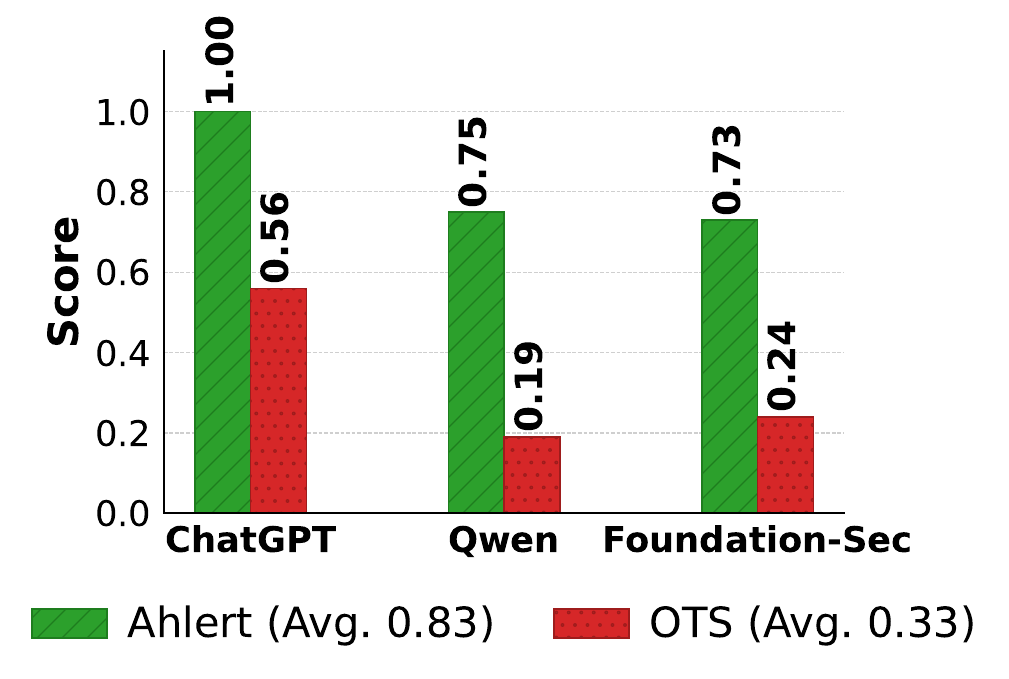}}
    \subfigure[F1 score.]{\includegraphics[width=0.32\linewidth]{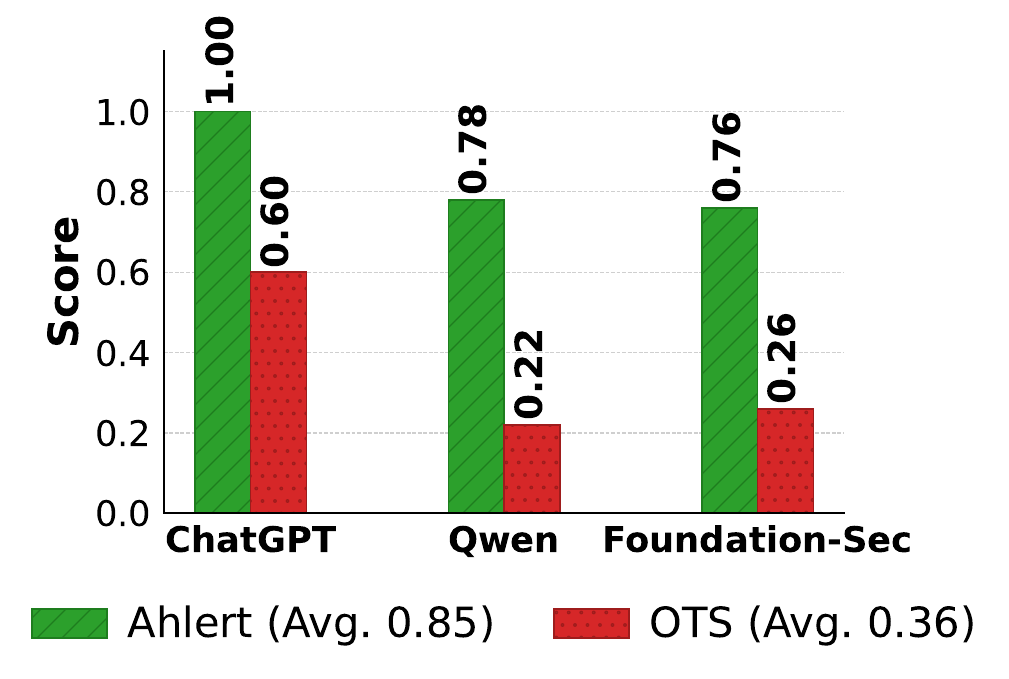}}
    \caption{\sol performance over APT41 reports.}
    \label{fig:eval-hist}
    \vspace{-0.3cm}
\end{figure}

Fig.~\ref{fig:eval-hist} reports the precision{\akash, the share of generated leads matching ground truth}, recall{\akash, the share of ground truth leads recovered}, and F1{\akash, their harmonic mean}, by each generator.
For every generator, 
\sol produces a consistent improvement over the \textsc{Ots} baselines. F1 rises from 0.22 to 0.78 for \blseven and from 0.26 to 0.76 for \bleight, with precision and recall improving in tandem (\eg for \blseven, precision from 0.26 to 0.81 and recall from 0.19 to 0.75). Averaged across the three generators, \sol raises mean precision from 0.40 to 0.87, recall from 0.33 to 0.83, and F1 from 0.36 to 0.85. The \blone, which serves as the reference model for ground-truth curation, attains a perfect F1 of 1.00 and is shown only as a completeness point rather than as an independent comparison.
This lift stems from the hybrid design: graph and entity expansion surface the implied ATT\&CK techniques and connected assets, and ontology grounding anchors each lead to a declared asset, yielding specific, verifiable hypotheses rather than the generic prose of single-route retrieval.

\begin{figure}[!t]
    \centering
    \includegraphics[width=\linewidth]{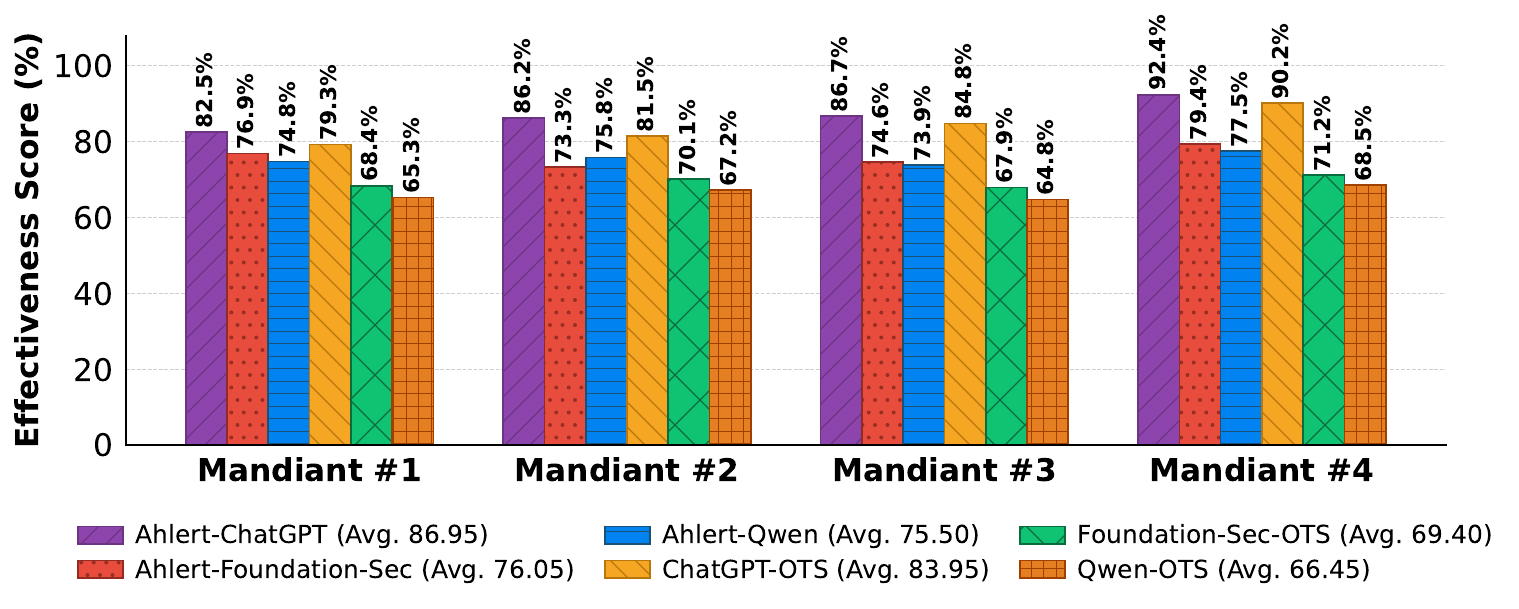}
    \caption{Lead relevance effectiveness score (\%).}
    \label{fig:rq1-effectiveness}
    \vspace{-0.3cm}
\end{figure}

We further computed the effectiveness score per system and report. This metric is calculated as the mean, taken over \sol's generated leads, of a weighted combination of three per-lead dimensions:
\begin{enumerate*}[(i)]
    \item \emph{Lead Relevance}: does the hypothesis relate to the reported threat activity?), 
    \item \emph{IoC Accuracy}: are named indicators correct and attributable to the report?, and
    \item \emph{Actionability}: is the lead specific enough to execute as a hunt query?.
\end{enumerate*}
Each dimension is scored on a 0 to 100 scale.
A high effectiveness score, therefore, indicates leads that are simultaneously on topic with the reported threat activity, accurate in their named indicators, and specific enough for an analyst to execute directly.
As shown in Fig.~\ref{fig:rq1-effectiveness}, \blone consistently achieves the highest effectiveness score across all four reports (avg.~86.95\%).
\bleight and \blseven follow at 76.05\% and 75.50\% respectively, with both benefiting measurably from the \sol pipeline relative to their off-the-shelf counterparts. Among the off-the-shelf systems, \blfour (avg.~83.95\%) is the strongest single baseline, reflecting GPT-4.1-mini's strong instruction-following even without retrieval; however, \blone still outperforms it by $3$~percentage points on average. \blnine and \blten, operating in pure off-the-shelf mode without a hybrid evidence context, score 66.45\% and 69.40\% respectively.
Overall, \sol improves not only precision and recall but also the practical usefulness of the leads, transforming generic LLM outputs into accurate, report-specific, and analyst-actionable hunt leads.

To further validate these results, we performed a dual validation protocol using both human and LLM-based assessment.
In the human-based validation, a domain expert evaluated each generated lead against its source report.
In the LLM-based validation, GPT-4o~\cite{openai2024gpt4} acted as an independent judge and scored each lead against the extracted ground truth.
Both validators assessed the same effectiveness dimensions, namely lead relevance, IoC accuracy, and actionability, and the resulting scores are reported in Table~\ref{tab:validation}.
Both validators confirm the same trend: \sol-augmented configuration scores higher than its \textsc{Ots} counterpart, averaged over the four reports, \blone leads \blfour (human 85.5 vs. 78.0; LLM 81.5 vs. 75.3), \blseven leads \blnine (68.0 vs. 65.0; 65.3 vs. 62.2), and \bleight leads \blten (71.5 vs. 68.6; 69.8 vs. 66.9).
\blone attains the highest overall scores (up to 94\% human and 90\% LLM on Mandiant~4). However, because this configuration is also used as the reference model for ground-truth curation, it is reported for completeness only. For the open-weight pairs, which are independent of the curation process, \sol gain is smaller but consistent across all four reports, about $+3$ percentage points for both \blseven over \blnine and \bleight over \blten under either validator.
The improvement brought by \sol is not an artifact of a single evaluation method. Human and GPT-4o validation agree closely on both the relative ranking and the magnitude of the gains, with the LLM judge being slightly more conservative by about 2-4 percentage points on average. This suggests that LLM-based validation can serve as a scalable proxy for expert assessment, while the human confirms the operational relevance of the generated hunt leads.

\begin{table}[!t]
    \centering
    \caption{Avg.\ results for human and LLM-based validation across threat reports, evaluated using different generators (\%).}
    \label{tab:validation}
    \begin{tabularx}{\linewidth}{
        >{\centering\arraybackslash}m{0.7cm}|
        >{\raggedright\arraybackslash}X
        >{\centering\arraybackslash}m{1.3cm}
        >{\centering\arraybackslash}m{1.3cm}
        }
        \hline
        \textbf{Report} & \textbf{Model} & \textbf{Human-based Validation} & \textbf{LLM-based Validation} \\ 
        \hline
        \multirow{6}{*}{\rotatebox[origin=c]{90}{Mandiant 1}}
         & \blone & \greencell{70.00} & \greencell{68.00} \\
         & \blfour & 66.00 & 62.00 \\
         & \blseven & 60.00 & 58.00 \\
         & \blnine & 56.30 & 54.30 \\
         & \bleight & 56.00 & 55.00 \\
         & \blten & 52.90 & 51.90 \\
        \hline
        \multirow{6}{*}{\rotatebox[origin=c]{90}{Mandiant 2}}
         & \blone & \greencell{94.00} & \greencell{88.00} \\
         & \blfour & 86.00 & 82.00 \\
         & \blseven & 66.00 & 60.00 \\
         & \blnine & 63.20 & 57.20 \\
         & \bleight & 80.00 & 78.00 \\
         & \blten & 77.30 & 75.30 \\
        \hline
        \multirow{6}{*}{\rotatebox[origin=c]{90}{Mandiant 3}}
         & \blone & \greencell{84.00} & \greencell{80.00} \\
         & \blfour & 74.00 & 72.00 \\
         & \blseven & 66.00 & 65.00 \\
         & \blnine & 62.90 & 61.90 \\
         & \bleight & 76.00 & 74.00 \\
         & \blten & 72.60 & 70.60 \\
        \hline
        \multirow{6}{*}{\rotatebox[origin=c]{90}{Mandiant 4}}
         & \blone & \greencell{94.00} & \greencell{90.00} \\
         & \blfour & 86.00 & 85.00 \\
         & \blseven & 80.00 & 78.00 \\
         & \blnine & 77.50 & 75.50 \\
         & \bleight & 74.00 & 72.00 \\
         & \blten & 71.70 & 69.70 \\
        \hline
    \end{tabularx}
\end{table}

\begin{figure}[!t]
    \centering
    \begin{minipage}[t]{0.49\linewidth}
        \centering
        \includegraphics[width=\linewidth]{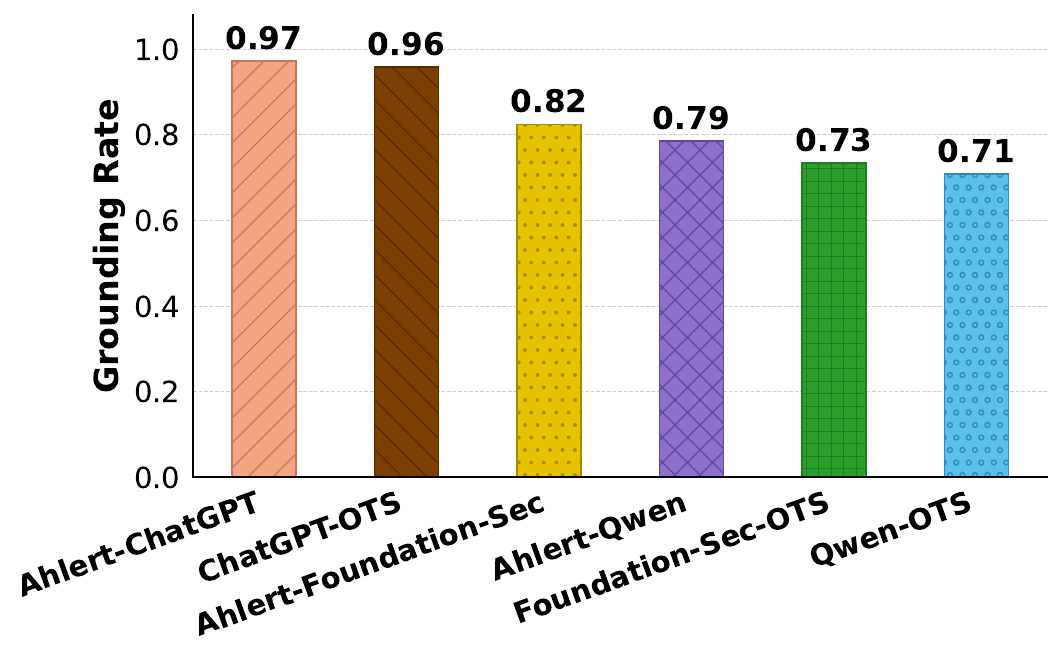}
        \caption{Grounding rate.}
        \label{plot:grounding}
    \end{minipage}%
    \hfill
    \begin{minipage}[t]{0.49\linewidth}
        \centering
        \includegraphics[width=\linewidth]{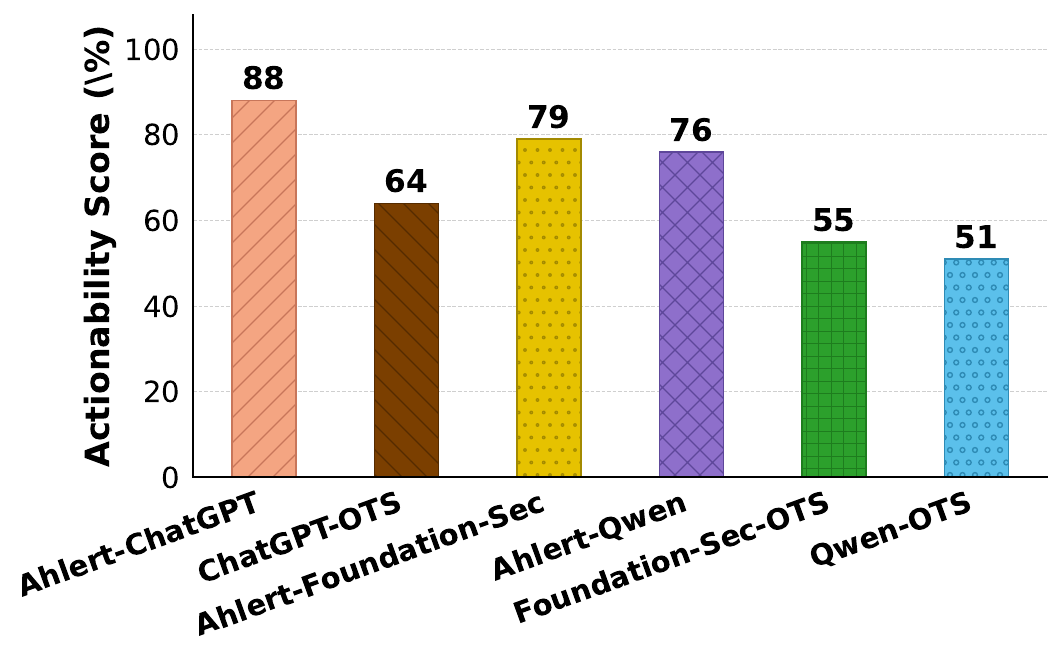}
        \caption{Actionability score.}
        \label{plot:actionability}
    \end{minipage}
    \vspace{-0.3cm}
\end{figure}

\smallskip
\noindent \textbf{RQ2 - \textit{Contextual Correctness}: Does \sol correctly ground threat to the enterprise network defined in the ontology?}
\noindent \textsc{RQ2} examines whether \sol correctly grounds threat behaviors to entities that exist in the defender's environment. For each hunt lead, we verify that at least one entity mentioned in the lead (\ie host name, service, IP address, or application) corresponds to an entity declared in the system ontology. 
We define the \emph{grounding rate} as the fraction of leads that pass this verification. A high grounding rate indicates that \sol effectively constrains generation to the defender's operational context rather than producing generic or hallucinated recommendations. 
As shown in Fig.~\ref{plot:grounding}, \blone achieves the highest rate (97.2\%), followed closely by \blfour (95.8\%). \sol-augmented open-weight models, \bleight and \blseven, reach 82.4\% and 78.6\%, respectively. \textsc{Ots} open-weight models, \blten and \blnine, achieve lower grounding rates of 73.4\% and 70.8\% respectively.
The gap reflects the ontology-constrained rewriting step: proprietary models reliably conform to the structured prompt mapping behaviors onto declared entities, whereas smaller open-weight models only partially follow it. \sol thus improves not only lead quality but also operational grounding

\smallskip
\noindent \textbf{RQ3 - \textit{Actionability}: Are the generated hunt leads sufficiently specific and structured for a threat hunter to act upon?}
\noindent RQ3 assesses whether the leads are structured and specific enough to execute without additional interpretation. We evaluate actionability along three dimensions: (i) whether the lead begins with a prescribed imperative verb (\eg \textit{search}, \textit{query}, \textit{scan});
(ii) whether it references at least one ATT\&CK technique or CVE; and (iii) whether the metrics block specifies concrete scoping parameters (\eg hosts, users, events).
Fig.~\ref{plot:actionability} reports the \emph{actionability score} for each configuration as the percentage of leads that satisfy all three criteria simultaneously.
\blone attains the highest actionability score (88\% of leads satisfy all three criteria), followed by \bleight at 79\% and \blseven at 76\%; all three reliably open with an imperative verb and embed an ATT\&CK identifier or CVE, as the ontology-grounded prompt enforces that structure.

The \textsc{Ots} systems trail substantially: \blfour reaches 64\%, while open-weight baselines \blten and \blnine reach only 55\% and 51\%. By dimension, (i) is met by nearly every system; (ii) is the primary differentiator, as \textsc{Ots} generators often describe behavior without a concrete identifier; and (iii) is most often missing, as \textsc{Ots} leads omit executable bounds.

\noindent \textbf{RQ4 - \textit{Ablation Study}: What is the contribution of retrieval, ontology grounding, and threshold to overall hunt lead quality?}

\begin{figure}[!t]
    \centering
    \begin{minipage}[t]{0.5\linewidth}
        \centering
        \includegraphics[width=\linewidth]{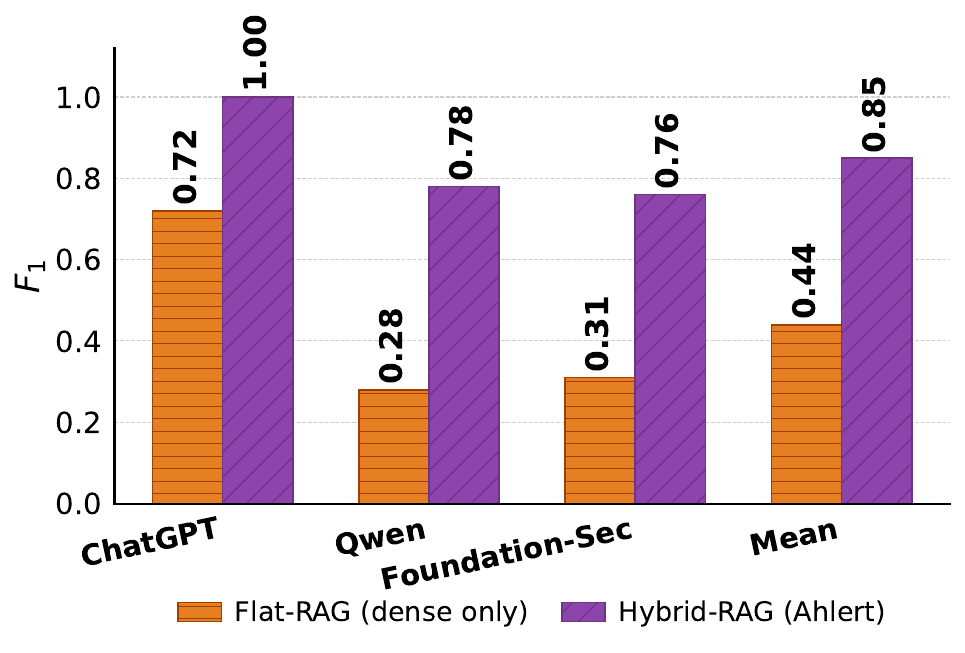}
        \caption{Impact of hybrid retrieval.}
        \label{plot:flat-vs-hybrid}
    \end{minipage}%
    \hfill
    \begin{minipage}[t]{0.5\linewidth}
        \centering
        \includegraphics[width=\linewidth]{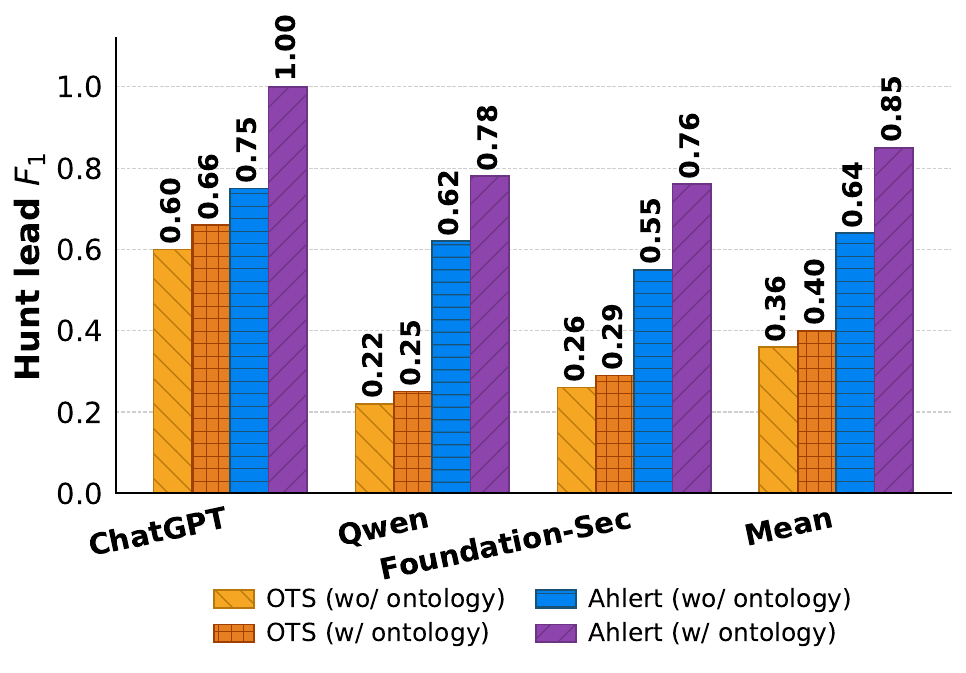}
        \caption{Impact of ontology.}
        \label{plot:impact-ontology}
    \end{minipage}%
    \hfill
    \begin{minipage}[t]{0.5\linewidth}
        \centering
        \includegraphics[width=\linewidth]{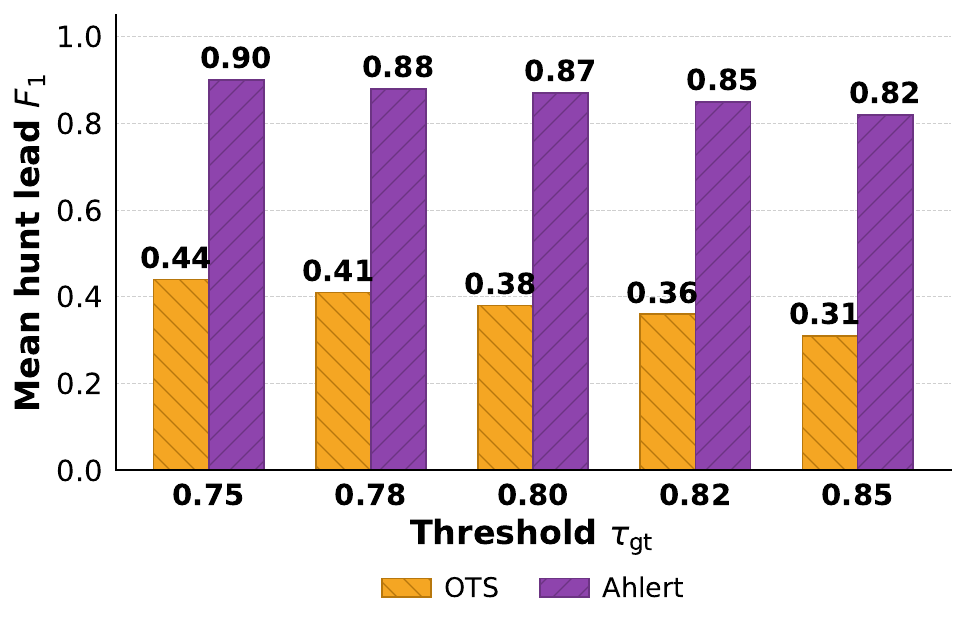}
        \caption{Impact of the threshold.}
        \label{plot:impact-threshold}
    \end{minipage}
    \vspace{-0.3cm}
\end{figure}

\subhead{$\bullet$ Impact of retrieval}
Fig.~\ref{plot:flat-vs-hybrid} isolates the retriever by comparing \sol with a single-route flat-RAG retriever (dense vector search only, graph and entity expansion disabled) against the full hybrid pipeline. Hybrid retrieval doubles the flat-RAG baseline, raising mean F1 from 0.44 to 0.85 ($\approx2\times$). For reference, the off-the-shelf generator without retrieval scores only 0.36 mean F1 (\textsc{Ots}, Fig.~\ref{fig:eval-hist}), so dense retrieval alone adds little over \textsc{Ots}; the decisive lift comes from graph expansion and ontology grounding.

\subhead{$\bullet$ Impact of ontology grounding}
Fig.~\ref{plot:impact-ontology} reports hunt lead F1 score across all three generators,
using different retrieval (\textsc{Ots} vs. hybrid graph retrieval) and with and without ontology grounding.
Isolating ontology grounding while holding the retriever fixed reveals a value strongly complementary to graph retrieval. On top of hybrid graph retrieval, it lifts mean F1 from 0.64 to 0.85 ($+0.21$), with per-generator gains of $+0.25$ for \blone (0.75$\rightarrow$1.00), $+0.16$ for \blseven (0.62$\rightarrow$0.78), and $+0.21$ for \bleight (0.55$\rightarrow$0.76). On top of the flat off-the-shelf retriever, the same grounding yields only $+0.04$ mean (0.36 to 0.40, at most $+0.06$ per generator), roughly $5\times$ less, since the hybrid retriever surfaces the concrete entities the ontology binds onto. The effect is largest for open-weight generators, narrowing much of the proprietary gap.

\subhead{$\bullet$ Impact of threshold}
Fig.~\ref{plot:impact-threshold} shows the mean hunt lead F1 score at different threshold $\tau_{\mathrm{gt}}$.
for \textsc{Ots} baseline and \sol-augmented systems.
Two patterns hold across the whole range.
\emph{First}, absolute F1 declines gradually as the threshold tightens (\sol 0.90 to 0.82, \textsc{Ots} 0.44 to 0.31), since a stricter cut-off admits fewer loosely related matches; the decline is smooth rather than abrupt, indicating genuine matches cluster well above the cut-off.
\emph{Second}, the ordering is invariant: \sol outperforms \textsc{Ots} at every threshold, and the margin \emph{widens} from $+0.46$ at $\tau_{\mathrm{gt}}=0.75$ to $+0.51$ at $\tau_{\mathrm{gt}}=0.85$, as generic \textsc{Ots} leads lose matches faster than ontology-grounded ones.
We adopt $\tau_{\mathrm{gt}}=0.82$ as it demands a close semantic paraphrase of an expert reference lead yet lies on the stable part of the curve where \sol scores 0.85.

\subhead{Discussion}
\sol does not surface unconstrained model output: its verification step discards any candidate lead unsupported by a retrieved evidence unit, suppressing fabricated techniques and indicators. Every actionable lead must cite an ATT\&CK technique or CVE drawn from the retrieved evidence (RQ3), and 70\%--97\% of leads reference assets that genuinely exist in the defender ontology (RQ2); a dedicated faithfulness benchmark is left to future work.

Despite these promising results of \sol, several limitations remain:
\begin{enumerate*}[(i)]
    \item the ground truth was curated starting from \blone output, so that configuration's F1 of 1.00 is a reference upper bound rather than an independent comparison;
    \item our evaluation centers on four reports for a single threat actor (APT41); \sol's pipeline is actor-agnostic, but cross-actor validation remains future work;
    \item the defender system ontology is a compact 79-instance proof of concept; \sol's retrieval and grounding logic is independent of ontology scale, but large-scale evaluation remains future work;
    \item lead quality depends on ontology completeness: when a relevant asset is absent from the snapshot, the grounding filter falls back to the unfiltered lead, which accounts for most residual ungrounded cases (RQ2);
    \item our automated relevance metric uses an SBERT matcher (all-MiniLM-L6-v2, $\tau_{\mathrm{gt}}=0.82$) as a proxy for human judgment, mitigated by parallel human validation but still a conservative design choice; and
    \item our evaluation does not yet include deployment in a live SOC, so operational scalability and analyst usability remain to be measured.
\end{enumerate*}

Future work will address these gaps through automated ontology expansion, harder domain/range generation constraints, and analyst feedback loops that continuously refine lead quality against expert scoring.
\section{Related Work}
\label{sec:related_work}

\subhead{Cyber Threat Intelligence Extraction}
A long line of work converts narrative threat reports into structured artifacts that downstream tooling can consume. Lekssays \etal\cite{lekssays2025azerg} introduce \textsc{Azerg}, a tool that fine-tunes general-purpose LLMs on four sequential subtasks.
Kim \etal\cite{kim2026anchor} follow a complementary path with \textsc{Anchor}, a schema-agnostic knowledge-graph construction system that pairs a search-and-navigate hybrid ontology-discovery mechanism with \textsc{SHACL}-based validation to assign schema-compliant types across large ontologies.
\textsc{Ctinexus}~\cite{cheng2025ctinexus} takes a third approach: rather than fine-tune, it uses in-context-learning prompts with optimal demonstration retrieval and hierarchical entity alignment to build a cybersecurity knowledge graph.
The recent SoK~\cite{buchel2025sok} systematizes more than 40 such efforts and finds that most stop at the entity or technique-mapping layer and rely on incompatible custom ontologies, leaving outputs hard to reuse and of limited value to a SOC analyst who must still hand-craft the hunt query.
\textit{In contrast}, \sol treats structured extraction as a \emph{means}: techniques and indicators feed the KB-backed graph queried at generation time, yielding a ranked, environment-consistent list of hypotheses rather than a static extraction bundle.

\subhead{RAG for Threat Intelligence}
RAG~\cite{lewis2020rag} is the default recipe for grounding LLM outputs in external knowledge, and several recent systems specialise it for cybersecurity~\cite{ahmadou2026automating}.
Lekssays \etal\cite{lekssays2025techniquerag} present \textsc{TechniqueRAG}, a retrieval-augmented~\cite{lewis2020rag} framework that maps free-form CTI text to adversarial ATT\&CK techniques by pairing instruction-tuned generation with retrieval over a technique corpus. 
Kurniawan \etal\cite{kurniawan2025agcyrag} push further with \textsc{AgCyRAG}, an agentic framework combining vector retrieval with Cypher and SPARQL traversals over a security knowledge graph. 
The general-purpose \textsc{GraphRAG} framework~\cite{edge2024graphrag} informs our own hybrid-context-assembly design.
A practical lesson is that hybridizing graph and vector evidence is necessary but not sufficient: the context must still become something an analyst can act on, yet existing CTI-RAG systems stop at technique annotation, free-form Q\&A (\eg~\cite{lekssays2025techniquerag, kurniawan2025agcyrag}), or graph construction (\eg~\cite{cheng2025ctinexus}).
\textit{In contrast}, \sol fuses semantic similarity, graph traversal~\cite{francis2018cypher}, and entity expansion into a unified evidence bundle grounded via a system ontology, yielding leads that are operationally actionable on the defender's estate.

\section{Conclusion}
\label{sec:conclusion}
This paper proposed \sol, a hybrid retrieval-and-generation system for automatically extracting actionable, evidence-grounded hunt leads from unstructured CTI reports. 
\sol \emph{unifies} flat retrieval across sources, graph traversal across knowledge, and controlled natural-language generation, and is \emph{environment-aware}: leads are filtered against a formal description of the defender's infrastructure so only operationally actionable hypotheses reach the analyst. 
Across four APT reports, the full hybrid pipeline raises mean F1 by $\approx2\times$ (0.44 to 0.85) over a single-route flat-RAG baseline, the \sol-augmented GPT attains the highest effectiveness score (avg.~86.95\%), and all three \sol-augmented generators exceed their off-the-shelf counterparts.

\section*{Acknowledgement}
This work was made possible in part through the support of the National Cybersecurity Consortium and the Government of Canada. It was also supported in part by Ericsson Research and the Security Research Centre of Concordia University.
The authors would like to thank \textit{Jan Willekens} and \textit{Jesus Alatorre} from Ericsson Cyber Defense Center for their invaluable feedback.

\section*{Model-Aware Prompting}
\label{app:prompt}
The example shows the prompt used by \sol to synthesize hunt leads from the retrieved evidence bundle.

\begin{tcolorbox}[
  colback=gray!5,
  colframe=black,
  coltitle=white,
  title=Example of a prompt,
  colbacktitle=black,
  fonttitle=\bfseries,
  arc=2pt,
  boxrule=0.5pt,
  left=4pt,
  right=4pt,
  top=3pt,
  bottom=3pt,
  float, floatplacement=!t 
]
{\scriptsize
\textbf{System:} You are a senior threat hunting analyst. Your task is to turn analyzed Cyber Threat Intelligence into concrete, investigable hunt leads that a SOC can execute against its own telemetry.

\textbf{User:} Read the evidence bundle and produce a ranked set of hunt leads, each grounded in the retrieved evidence and the defender's declared environment.

\textbf{Input:} \textless \textit{evidence bundle}\textgreater~ matched CTI passages, their MITRE ATT\&CK techniques, knowledge graph context, and system-ontology entities defining the defender's environment.

\textbf{Constraints:} Each lead must reference at least one ontology entity and at least one ATT\&CK technique or CVE, begin with an imperative verb, and carry explicit scoping parameters.

\textbf{Instructions:} Emit each lead as a structured object with \textit{summary}, \textit{severity}, \textit{priority}, \textit{impact}, and a \textit{metrics} block (hosts, users, events, time window). Discard candidates that lack supporting evidence or are inconsistent with the environment, merge near-duplicates, and rank the remainder by calibrated confidence.\par
}
\end{tcolorbox}

\bibliographystyle{IEEEtran}
\bibliography{Ref}

\end{document}